\documentclass[aps,prl,twocolumn,superscriptaddress,showpacs]{revtex4-1}
\usepackage{graphicx,amsmath}
\usepackage{amssymb}

\begin{document}
\title{Orbital selective phase transition induced by different magnetic states:\\ A dynamical cluster approximation study}

\author{Hunpyo Lee}
\affiliation{Institut f\"ur Theoretische Physik, Goethe-Universit\"at Frankfurt, Max-von-Laue-Stra{\ss}e 1, 60438 Frankfurt am Main, Germany}
\author{Yu-Zhong Zhang}
\affiliation{Institut f\"ur Theoretische Physik, Goethe-Universit\"at Frankfurt, Max-von-Laue-Stra{\ss}e 1, 60438 Frankfurt am Main, Germany}
\affiliation{Department of Physics, Tongji University, Shanghai, 200092 P. R. China}
\author{Harald O. Jeschke}
\affiliation{Institut f\"ur Theoretische Physik, Goethe-Universit\"at Frankfurt, Max-von-Laue-Stra{\ss}e 1, 60438 Frankfurt am Main, Germany}
\author{Roser Valent\'\i}
\affiliation{Institut f\"ur Theoretische Physik, Goethe-Universit\"at Frankfurt, Max-von-Laue-Stra{\ss}e 1, 60438 Frankfurt am Main, Germany}

\date{\today}

\begin{abstract}

Motivated by the unexplored complexity of phases present in the
multiorbital Hubbard model, we analyze in this work the behavior of a
degenerate two-orbital anisotropic Hubbard model at half filling where
both orbitals have equal bandwidths and one orbital is constrained to
be paramagnetic (PM), while the second one is allowed to have an
antiferromagnetic (AF) solution.  Such a model may be relevant for a
large class of correlated materials with competing magnetic states in
different orbitals like the recently discovered Fe-based
superconductors.  Using a dynamical cluster approximation we observe 
that novel orbital selective phase transitions
 appear regardless of the strength of the Ising Hund's rule coupling
 $J_z$. Moreover, the PM orbital undergoes a transition from a Fermi
 liquid (FL) to a Mott insulator through a non-FL phase
 while the AF orbital shows a transition from a FL to an AF insulator
 through an AF metallic phase.  We discuss the
 implications of the results in the context of the Fe-based
 superconductors.
\end{abstract}

\maketitle

The existence of an orbital selective phase transition (OSPT) induced
by the interplay of a narrow band of localized electrons and a wide
band of itinerant electrons has evoked considerable interest since the
discovery of exotic phase transitions in
Ca$_{2-x}$Sr$_x$RuO$_4$~\cite{Nakatsuji2000,Anisimov2002}.  Presently,
OSPT are being intensively investigated as an alternative to
conventional Mott transitions happening simultaneously in all orbitals
in correlated
systems~\cite{Koga2004,Knecht05,Biermann2005,Liebsch2005,Bouadim09,Lee2010,Lee2010a,Werner2007,Medici2009}.
Previous works on OSPT were either based on degenerate orbitals with
inequivalent bandwidths at integer
filling~\cite{Koga2004,Knecht05,Biermann2005,Liebsch2005,Bouadim09,Lee2010,Lee2010a}
or focused on orbitals with degeneracies lifted completely or
partially by the crystal field splitting at different
fillings~\cite{Werner2007,Medici2009}.  Recently a new class of
multiorbital systems -the Fe-based superconductors- where
superconductivity appears to be strongly related to magnetism has
opened a new field of debate. In the present work, we investigate a
minimal two-band Hubbard model of relevance to understand the nature
of the antiferromagnetic metallic state and the experimentally
observed reduced magnetic moment in Fe-based superconductors.  We will
show that inclusion of (short-range) spatial fluctuations which are
present in the dynamical cluster approximation (DCA) allows also for a
possible non-Fermi liquid phase. Such a phase has been recently
discussed for BaFe$_2$(As$_{1-x}$P$_x$)$_2$~\cite{Kasahara2010} and
FeSe~\cite{Aichhorn2010}.  Our results suggest also a new mechanism
for an OSPT induced by different magnetic states in different
orbitals.

We consider
a degenerate two-orbital anisotropic Hubbard model at half filling
where both orbitals have equal bandwidths and one orbital is
constrained to be paramagnetic (PM) (PM orbital), while the
second one is allowed to have an antiferromagnetic (AF) solution (AF orbital).
This model should capture  the coexistence  of frustrated
bands where magnetic order is suppressed  with unfrustrated
bands which can order antiferromagnetically
due to perfect nesting,
as observed in bandstructure calculations~\cite{Miyake2010}. 
We employ the DCA~\cite{Maier2005} with a cluster size of $N_c=4$ in combination
with a weak-coupling continuous time quantum Monte Carlo 
algorithm~\cite{Rubtsov2005}. Such
an approach ensures the consideration of the dynamical fluctuations
and the inclusion of a symmetry-breaking state as well as  spatial
fluctuations which cannot be captured in most of  previous OSPT works using a
single-site dynamical mean field theory
(DMFT)~\cite{Koga2004,Knecht05,Biermann2005,Liebsch2005,Werner2007,Medici2009}.

\begin{figure}
\includegraphics[width=0.33\textwidth,angle=270]{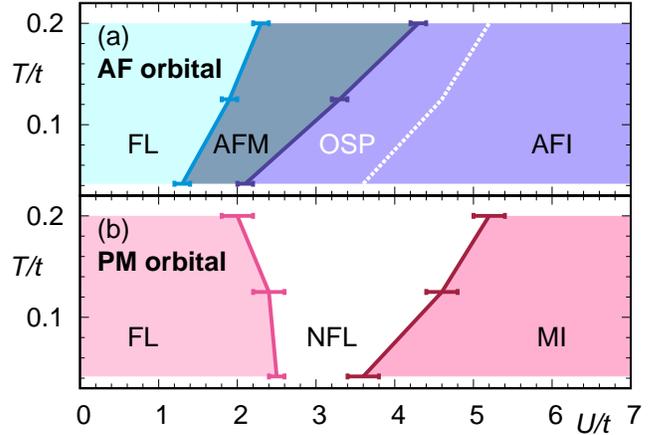}
\caption{(Color online) The $U-T$ phase diagram of (a) AF and (b) PM orbitals for $J_z=U/8$ and $U'=3U/4$.
 The non-FL and AF metal are determined by the self-energy neither showing divergent behavior nor approaching zero as the Matsubara frequency goes to zero and the spin-dependent symmetry breaking field remaining finite in the metallic state, respectively.} \label{fig:diagram}
\end{figure}

We find a novel OSPT: In the PM orbital, a phase transition from a
Fermi liquid (FL) to a non-FL state followed by a Mott
metal-to-insulator transition (MIT) is detected, while in the AF
orbital two phase transitions from a PM metal to an AF insulator
through an AF metallic state are observed, regardless of the strength
of the Ising-type Hund's rule coupling $J_z$.  All the transitions
occur at different critical values of $U/t$. Note that here
 the OSPT occurs when
 the PM orbital undergoes a Mott MIT while the AF orbital remains an
 AF insulator. Furthermore, we also discuss  the nature of the
Mott MIT in the PM orbital at $J_z=0$ where the ferromagnetic
interaction between PM and AF orbitals vanishes.  Such a consideration
simultaneously allows for  the study of (i) strong orbital fluctuations between PM
and AF orbitals as well as
 (ii) the effect that the AF order in the AF orbital induces on the PM
orbital. Moreover, comparable to the single-orbital Hubbard model within a
small cluster~\cite{Park2008,Gull2008,Zhang2007}, the cooperation
between the spatial AF order due to perfect nesting and frustration
induced by short-range spatial fluctuations remains in our model.
Nevertheless, while the single-orbital Hubbard model shows a
first-order Mott MIT~\cite{Park2008,Gull2008}, we observe a continuous
Mott MIT in the PM orbital with a critical interaction $U_c/t=4.0$
close to that obtained for the single-orbital system by the same DCA
method with a cluster size of
$N_c=4$~\cite{Gull2008,Lin2008}. Finally, we discuss the implications
of the non-Fermi liquid phase as well as the small magnetic moment in
the AF orbital in the context of the Fe-based superconductors.

The degenerate two-orbital anisotropic Hubbard model with a
PM and an AF orbital at
half-filling on the square lattice can
be written as
\begin{eqnarray} \label{hamil}
  H &=&-\sum_{\langle ij\rangle m\sigma}  t_m c^{\dagger}_{im\sigma}
  c_{jm\sigma}+\,U\sum_{im}n_{im\uparrow} n_{im\downarrow} \nonumber \\
  &&+\sum_{i\sigma\sigma'}(U'-\delta_{\sigma \sigma'} J_z)n_{i1\sigma}
  n_{i2\sigma'},
\end{eqnarray}
where $c_{im\sigma}(c^{\dag}_{im\sigma})$ is the annihilation
(creation) operator of an electron with spin $\sigma$ at the $i$-th
site with orbital index $m=(1,2)$. $t_m$ is the hopping matrix element
between site $i$ and $j$, $U$ and $U'$ are intra-orbital and
inter-orbital Coulomb repulsion integrals, respectively, and $J_z
n_{i1 \sigma}n_{i2 \sigma}$ is the Ising-type Hund's rule coupling
term~\cite{comment_Hund}. The bandwidth is $W=8$ ($t=1$). 
  In order to simulate our
anisotropic Hubbard model with a PM and an AF orbital, a
spin-dependent symmetry breaking field is applied on the AF orbital in
the first iteration, while we always keep the PM solution under the
condition of $\overline{G}_{PM}(i\omega_n) =
\frac{1}{2}(\overline{G}_{\uparrow}(i\omega_n)+\overline{G}_{\downarrow}(i\omega_n)$ for the
PM orbital at each iteration. The converged Green's functions are obtained from the DCA and Dyson equations.

First, we would like to discuss the phase diagram for $J_z=U/8$ and
$U'=3U/4$ in the $U-T$ plane shown in Fig.~\ref{fig:diagram}.
Analogous to the phase diagram of the single-orbital Hubbard model
with a PM solution obtained from cluster-DMFT with
$N_c=4$~\cite{Park2008}, we observe that the FL, non-FL and Mott
insulator phases are present in the PM orbital. The metallic regions are
shrunk due to the enhancement of spatial AF correlations with
decreasing temperature.  The main difference between single-orbital
and our two-orbital Hubbard model is that the first-order MIT
 present in the single-orbital Hubbard model, is replaced by a
continuous MIT in the two-orbital Hubbard model (see
Fig.~\ref{fig:diagram}~(b)).  The coupling between the PM orbital with
spatial AF correlations and the AF orbital with AF order suppresses
the first-order transition. Furthermore, in the weak-coupling region,
while the AF insulator is present in the cluster-DMFT calculation of
the single-orbital Hubbard model~\cite{Fuhrmann2007}, in our AF orbital the AF symmetry
breaking field is completely suppressed due to  thermal and orbital
fluctuations between PM and AF orbitals (see
Fig.~\ref{fig:diagram}~(a)). In fact, in the intermediate region the
AF orbital shows an AF metallic phase and as the interaction $U/t$ is
increased, the AF insulator state is reached. The orbital selective
phase where a metallic state in the PM orbital and an insulating state
in the AF orbital coexist, is clearly observed in the intermediate
regime. Such a phase is induced by different magnetic states in the
two orbitals. This mechanism is distinct from previously proposed
ones, such as change of filling in non-degenerate or partially
degenerate orbitals~\cite{Werner2007,Medici2009}, variation of
bandwidth in different
orbitals~\cite{Koga2004,Knecht05,Biermann2005,Liebsch2005,Lee2010,Bouadim09}
or differences of frustration strength in different
orbitals~\cite{Lee2010a}.

\begin{figure}
\includegraphics[width=0.39\textwidth,angle=270]{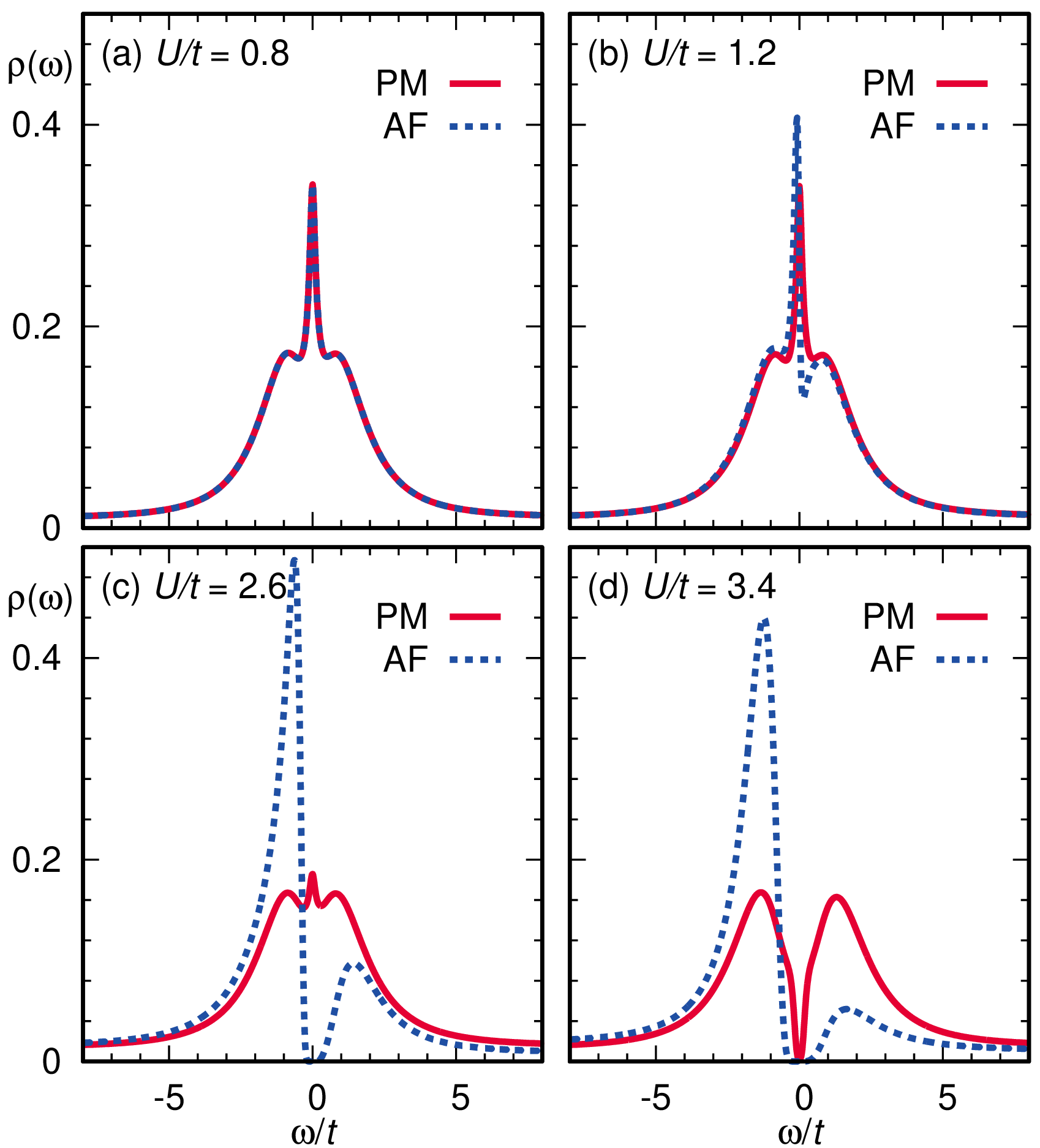}
\caption{(Color online) Density of states $\rho(\omega)$ per site for different values of the interaction
  strength $U/t$, with $U'=U/2$ and $J_z=U/4$ at
  $T/t=1/24$. Here, only the $\rho(\omega)$ for spin up on one sublattice is shown when the AF orbital is in an AF state.} \label{fig:DOS}
\end{figure}

In the following, we will analyze the various states in our
phase diagram.
 Figs.~\ref{fig:DOS} (a)-(d) show the density of states (DOS)
for the two orbitals for several values of the
interaction strengths $U/t$ with $U'=U/2$ and $J_z=U/4$ at
$T/t=1/24$, obtained using the maximum entropy method.
In the weak-coupling region $(U/t=0.8)$ both the PM and the AF orbital
are in the PM metallic phase.
 This
 results in
the same DOS for both orbitals including a quasiparticle peak at the Fermi level 
 as shown in Fig.~\ref{fig:DOS}~(a). As the interaction
increases ($U/t=1.2$), the AF orbital - while remaining metallic -
starts to display AF order
 as shown in Fig.~\ref{fig:DOS}~(b). In this case,
 the spin up DOS is pairwise equal to the spin down  DOS
  on each site.
  The PM orbital remains 
 in the PM metallic phase
 with a quasiparticle peak at the Fermi level. In
the intermediate region at $U/t=2.6$ (Fig.~\ref{fig:DOS}~(c))
 the metallic PM orbital coexists with the
insulating  AF orbital. When the interaction is further increased
to the strong-coupling region $(U/t=3.4)$, insulating states appear in
both orbitals as displayed in Fig.~\ref{fig:DOS}~(d).  Note that we
found the OSPT behavior for all investigated values of $J_z$.

\begin{figure}
\includegraphics[width=0.35\textwidth,angle=270]{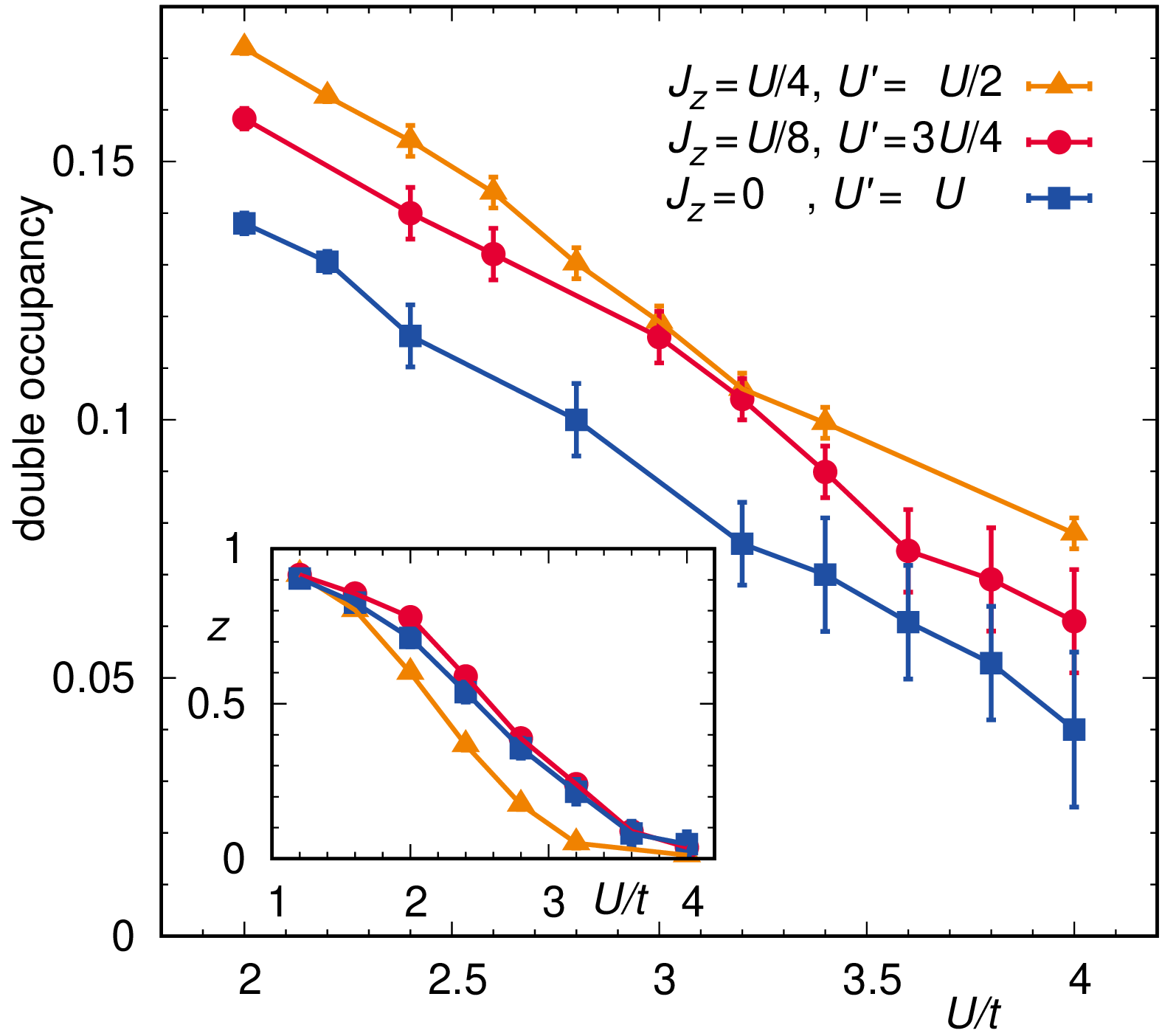}
\caption{(Color online) The double occupancy of the PM orbital as a function of $U/t$
  for three values of $U'$ and $J_z$ at
  $T/t=1/24$, with error bars as indicated.
Inset:  Quasiparticle weight of the PM orbital s a function of $U/t$.} \label{fig:doubleoccupancy}
\end{figure}

Next, we would like to clarify the nature of the Mott MIT in the PM
orbital.  In the case of the single-orbital Hubbard model, the
double occupancy shows hysteresis with coexistence regions
indicating a first-order phase transition~\cite{Park2008}.  Since our
two-orbital system allows for additional orbital degrees of freedom,
it is interesting to check whether the nature of the transition is
changed due to the coupling between AF and PM orbitals.  In
Fig.~\ref{fig:doubleoccupancy}, we plot the double occupancy as a
function of $U/t$ with different $U'$ and $J_z$ at $T/t=1/24$. While
 $U_c/t$ increases from $3.0$ to $3.5$ and $3.6$ when
$J_z$ is decreased from $U/4$ to $U/8$ and to $0$ with the constraint
of $U=U'+2J_z$, in none of the cases   a signature of
hysteresis in the double occupancy is found.  The disappearance of the
first-order transition is attributed to (i) the interaction between the
orbitals and (ii) the cooperation of spatial AF correlations in the PM
orbital and AF order in the AF orbital at all values of $J_z$.
In the inset of Fig.~\ref{fig:doubleoccupancy} we also show the evolution of
the spectral weight for the PM orbital.

\begin{figure}
\includegraphics[width=0.35\textwidth,angle=270]{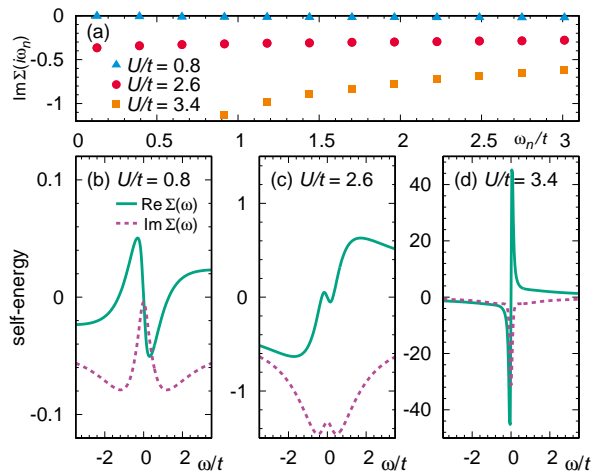}
\caption{(Color online) Self-energies at the Fermi surface ${\bf K}=(\pi,0)$
  as a function of (a) Matsubara frequency and of (b)-(d) real frequency $\omega /t$  at (b)  $U/t=0.8$,
(c) $U/t=2.6$
   and (d) $U/t=3.4$ for $U'=U/2$, $J_z=U/4$ at $T/t=1/24$ on the PM
  orbital.}
\label{fig:selfenergy}
\end{figure}

Recently, the non-FL behavior in the multi-orbital as well as in the
single-orbital Hubbard model has been extensively studied within
DMFT or its cluster
extension~\cite{Biermann2005,Medici2009,Lee2010,Ishida2010,Park2008,Gull2008,Zhang2007,Werner2008,Costi2007}.
Also, non-FL behavior has been observed in the multiorbital Fe-based
superconductors~\cite{Kasahara2010}.
 Using the maximum entropy method~\cite{Wang2009} 
for the analytical continuation of the
self-energy, we analyze the self-energy $\Sigma (\omega)$ at the Fermi
surface ${\bf K}=(\pi,0)$ of the PM orbital as a function of the real
frequency $\omega$ (see Figs.~\ref{fig:selfenergy}~(b)-(d)) for $U'=U/2$, 
$J_z=U/4$ at $T/t=1/24$. Im $\Sigma (i\omega_n)$ as a function
of  Matsubara frequency
is also shown in  Fig.~\ref{fig:selfenergy}~(a)
In Fig.~\ref{fig:selfenergy}~(b), $\text{Im} \Sigma (\omega = 0)$
approaches zero indicating a FL behavior.  Upon increasing the
interaction in Fig.~\ref{fig:selfenergy}~(c), the system still shows
metallic behavior but $\text{Im} \Sigma (\omega = 0)$ has a finite
value (see also Fig.~\ref{fig:selfenergy}~(a)).
 This suggests a non-FL behavior~\cite{comment}. In the strong-coupling
region ($U/t=3.4$), the self-energy $\text{Im} \Sigma (\omega=0)$
diverges (Fig.~\ref{fig:selfenergy}~(d)) and the system is in a Mott
insulating state. Also shown in Figs.~\ref{fig:selfenergy}~(b)-(d) is
the $\text{Re} \Sigma (\omega)$ obtained from the Kramers-Kronig
relation.  From these results we infer that the gap opening in the PM
orbital is caused by a blocking of the electron delocalization due to
the Coulomb repulsion (Mott insulator), while in the AF orbital the
quasi-particle peak is split into two parts above and below the Fermi
level due to the antiferromagnetism with a gap of $\Delta \approx mU$
(AF insulator), where $m$ is the magnetization and $U$ is the
interaction strength.  Here we would like to stress that the gap
opening in the AF orbital -- which favors antiferromagnetism --
happens at a lower interaction strength than it happens in the PM
orbital. Therefore, a Mott MIT occurs in the PM orbital while the AF
orbital remains an AF insulator.  As mentioned above, this indicates a
novel OSPT driven by different magnetic states in different orbitals.

\begin{figure}
\includegraphics[width=0.35\textwidth,angle=270]{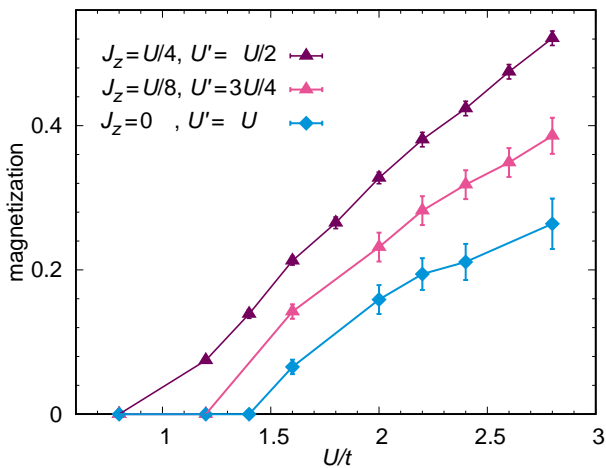}
\caption{(Color online) The staggered magnetization of the AF orbital as a function
  of $U/t$ for three values of $U'$ and
  $J_z$.} \label{fig:magnetization}
\end{figure}

Let us now analyze the AF metallic phase in the AF orbital at
intermediate couplings. Unlike the high-$T_c$ cuprates with an AF
insulating state in the normal phase, the undoped Fe-based
superconductors show AF metallic behavior with a small ordered
magnetic moment.  DFT
calculations~\cite{Mazin2008,Opahle2009,Zhang2009} overestimate the
values of the ordered magnetic moment compared to those obtained from
experiments.  With the consideration of an artificially negative $U$
in the DFT calculations, the small magnetic moment is
recovered~\cite{Ferber2010,Li2010}.  Nevertheless the mechanism for
the reduced ordered magnetic moment still remains
controversial~\cite{Si2008,Zhang2010,Bascones2010,Yang2010,Misawa2010,Yin2010}. Very
recently, the authors proposed a mechanism of coupling between
frustrated and unfrustrated orbitals within a two-orbital Hubbard
model, solved by single-site DMFT~\cite{Lee2010a} as a possible
mechanism for the reduced magnetic moment and the AF metallic state.
In the present work we study a related model (Eq.~\eqref{hamil}) by
DCA where spatial fluctuations -- absent in the single-site DMFT
calculation -- are partially considered. In the present model, keeping
one orbital always in the PM state can be viewed as an effective
frustration due to the suppression of long-range correlations by the
small cluster size~\cite{Park2008}.  In Fig.~\ref{fig:magnetization}
we present the staggered magnetization in the AF orbital as a function
of $U/t$ with different $U'$ and $J_z$ for $T/t=1/24$.
  We observe that, as the
interaction strength is increased, the staggered magnetization
increases continuously for all values of $J_z$ and remains small in
the range $U/t = 1.5 -2.5$ relevant for the Fe-based superconductors.
  This is also consistent with DMFT
results~\cite{Lee2010a}.

In conclusion, we have studied the two-orbital Hubbard model at half
filling where one orbital is constrained to be in the PM state and the
second orbital is allowed to have an AF solution.  Analysis of double
occupancy, magnetization and DOS shows that as $U/t$ increases, while
a continuous Mott MIT through non-FL behavior is observed in the PM
orbital, two phase transitions (PM metal to AF metal and AF metal to
AF insulator) are detected in the AF orbital.  We find a novel OSPT where a MIT occurs in the PM orbital while
the AF orbital remains an AF insulator. We ascribe the OSPT detected
in this work to the different magnetic states in the different
orbitals since other effects such as different bandwidths, crystal
field splitting and change of band filling have been avoided. Even
with inclusion of spatial fluctuations, the AF metal still survives in
the AF orbital in a certain range of interaction strengths. These
results support the scenario of a reduced ordered magnetic moment and
the existence of an AF metallic state in the Fe pnictides driven by
coupling between frustrated and unfrustrated
orbitals~\cite{Lee2010a}. Finally, by investigating the self-energy as
a function of Matsubara and real  frequency $\omega/t$, we identify an interaction
regime where non-FL behavior can be observed in the PM orbital.
This could be of  relevance  for  BaFe$_2$(As$_{1-x}$P$_x$)$_2$~\cite{Kasahara2010}
and  FeSe where non-FL behavior has been
discussed~\cite{Aichhorn2010}. 

{\bf Acknowledgments.-} We would like to thank the Deutsche
Forschungsgemeinschaft for financial support through grant FOR 1346
and the Helmholtz Association for support through HA216/EMMI.


\begin{thebibliography}{10}
\bibitem{comment_Hund} In the metallic regions -which are at the focus
  of the present study- a similar description is expected for Ising
  and rotationally invariant Hund couplings.
\bibitem{comment} We confirmed that $\text{Im} \Sigma (\omega = 0)$
  remains finite even at $T/t= 0.02$, giving us confidence in the
  proposed non-FL behavior.
\end{thebibliography}
\end{document}